\documentclass[preprint,12pt]{elsarticle}
\usepackage[utf8]{inputenc}
\usepackage[T1]{fontenc}
\usepackage{amsfonts}
\usepackage{mathrsfs,amsmath}
\usepackage{here}
\usepackage{graphicx}
\usepackage{algorithm}
\usepackage{color}
\usepackage{program}
\usepackage{subcaption}
\usepackage{fullpage}
\usepackage{hyperref}
\usepackage{booktabs,array}

\begin{document}

\begin{frontmatter}

\title{A semi-Lagrangian discontinuous Galerkin method for drift-kinetic simulations on GPUs}

\author[a]{Lukas Einkemmer}
\author[a]{Alexander Moriggl}

\address[a]{Department of Mathematics, University of Innsbruck, Austria}

\begin{abstract}
In this paper, we demonstrate the efficiency of using semi-Lagrangian discontinuous Galerkin methods to
solve the drift-kinetic equation using graphic processing units (GPUs). In this setting we propose a second order splitting scheme and a 2d semi-Lagrangian scheme in the poloidal plane. The resulting method is able to conserve mass up to machine precision, allows us to take large time steps due to the absence of a CFL condition and provides local data dependency which is essential to obtain good performance on state-of-the-art high-performance computing systems. We report simulations of a drift-kinetic ion temperature gradient (ITG) instability and show that our implementation achieves a performance of up to 600 GB/s on an A100 GPU.

\vspace{0.3cm}

\end{abstract}

\end{frontmatter}

\section{Introduction}
Kinetic equations have proven to be excellent models to understand the behavior of plasma. For plasmas with strong magnetic fields (such as in tokamaks), the full 6 dimensional Vlasov--Poisson or Vlasov--Maxwell simulation can be reduced to the 5d gyrokinetic (3d in space and 2d in velocity), see e.g.~\cite{lee1983gyrokinetic}, or the 4d drift-kinetic equations (3d in space and 1d in velocity), see e.g.~\cite{GRANDGIRARD2006395}. 

Semi-Lagrangian schemes~\cite{sonnendrucker1999semi} are in many cases the preferred choice for solving kinetic equations since these methods do not suffer from the same stability limitations as Eulerian methods (finite differences, finite volumes), see e.g.~\cite{filbet2003}. Moreover, these schemes allow, in general, more accurate results, assuming one is willing to pay the computational cost, than particle in cell methods (PIC), which introduce numerical noise especially in low density regions, see e.g. \cite{verboncoeur2005particle}. 

Often semi-Lagrangian methods are combined with splitting schemes~\cite{casas2017high,crouseilles2015hamiltonian,einkemmer2014convergence_a,einkemmer2014convergence_b}, which allows us to reduce the complexity of the full nonlinear equations to a sequence of usually linear lower dimensional problems. This permits us to design and use highly efficient methods that are tailored to these simpler sub-problems. It should be noted that for a large class of equations, for example  the Vlasov--Poisson model, the problem can be reduced to a sequence of one dimensional advective equations. However, this is not the case for the drift-kinetic model, where a 2d splitting substep has to be retained in order to resolve the characteristic in the poloidal (perpendicular to the magnetic field lines) plane accurately without introducing a large splitting error, see e.g.~\cite{GRANDGIRARD2006395, Cai2020comparison}.

The main idea of semi-Lagrangian schemes is to follow the characteristics backward or forward in time. Since the foot of the characteristics usually does not coincide with a grid point, an interpolation technique has to be applied, see e.g.~\cite{sonnendrucker1999semi}.  
Among the various interpolation techniques, cubic splines are often the favored option because they provide accurate results and introduce very little numerical diffusion \cite{filbet2003}. However, since in the high-dimensional setting of kinetic equations often large supercomputers are required, global methods such as spline interpolation impose significant difficulties to achieve good performance~\cite{Einkemmer20194d}. This is an even more serious issue on GPUs which are currently used extensively in high-performance computing (HPC). Consequently, local methods such as the semi-Lagrangian discontinuous Galerkin (SLDG) schemes have proven to be advantageous on GPUs~\cite{Einkemmer2020GPUs,Einkemmer2022}. In fact, we are not aware of a standard spline construction (i.e.~by solving tridiagonal linear systems) for semi-Lagrangian methods that has been implemented on GPUs. However, let us mention that local spline methods, as in~\cite{LatuSpline2010}, or high-order finite difference approximations of the required derivatives for constructing splines, as in~\cite{ROCHAFILHO2013}, have been implemented on GPUs. In \cite{mehrenberger2013vlasov} spline based interpolation is foregone in favor of the more easily parallelizable but more diffusive polynomial interpolation. In \cite{kormann2019massively} a massively parallel implementation (on CPU hardware) also uses polynomial interpolation. 

To tackle the 2d advective sub-problem in the splitting scheme, often a so-called backward semi-Lagrangian (BSL) scheme is applied where in the reconstruction step a cubic spline is used~\cite{GRANDGIRARD2006395}. This method does not conserve the mass of the system, which might lead to unphysical long time simulation results. Recently, a 2d semi-Lagrangian discontinuous Galerkin method has been developed in~\cite{Cai2021,Cai2017,Cai2019}. This method, besides conserving the mass exactly up to machine precision, is also local in contrast to the BSL scheme. Currently, to the best of our knowledge, no GPU implementation of this method exists, and it is not clear in advance if a reasonable performance on GPUs is possible. In this paper, we demonstrate that this goal can be achieved. Note that in~\cite{Crouseilles2014} a conservative scheme for the 2d problem has already been introduced. However, although better conservation properties than with the BSL method are achieved, mass is still not conserved up to machine precision. Hence, our implementation is the first splitting based scheme that achieves mass conservation up to machine precision. Moreover, a first order splitting scheme is used in~\cite{Crouseilles2014}, which we will extend to second order in this paper.

In most applications, the only metric of interest in HPC for high-dimensional kinetic equations in plasma physics is weak scaling. That is, problem size and computing resources are increased simultaneously and the execution time should not rise drastically. 
This is crucial since small-scale structures appear in plasma (this phenomenon is called filamentation) which can not be captured with low resolutions. Therefore, the size of the problem typically scales with the number of available resources.
The goal of this paper is to demonstrate efficient single-node performance, which is the first step in developing an efficient and scalable code.
Hence, the code used in this work, which is available at \texttt{https://bitbucket.org/leinkemmer/sldg}, can run both on multi-core shared memory CPU-based systems, and on single GPUs. We mainly use an A100 GPU and a dual-socket Intel Xeon Gold 6226R CPU, where we exploit all computational and memory resources of the underlying systems. We use OpenMP for the parallelization on the two CPUs and CUDA to parallelize on the GPU. This allows us to run simulations with a moderate resolution of approximately 200 grid points in each dimension on both platforms. 
For higher resolutions or an extension of the code to the 5d gyrokinetic equations~\cite{lee1983gyrokinetic}, a (cuda-aware) MPI implementation has to be considered in order to deal with the increasing memory requirements. This will be considered as future work.

This article is organized as follows. At first, we introduce the drift-kinetic model, then we describe the time splitting method and the numerical methods used in the splitting steps. Finally, we report the results of the simulation and analyze the performance of our code.

\section{Problem description}
We are interested in solving the 4d drift-kinetic equation in cylindrical coordinates~\cite{GRANDGIRARD2006395},

\begin{equation}
    \partial_tf - \frac{\partial_\theta \phi}{r}\partial_rf +   \frac{\partial_r \phi}{r}\partial_\theta f + v\partial_zf - \partial_z\phi\partial_vf=0
    \label{eq:dk}
\end{equation}
on the domain $(r,\theta,z,v) \in \Omega \times [0,L]\times [-v_{\text{max}},v_{\text{max}}]$, with $\Omega = [r_{\text{min}},r_{\text{max}}]\times[0,2\pi]$, and 
where $\phi=\phi(r,\theta,z)$ solves the quasi neutrality equation, 
\begin{equation}
    -\left[\partial_r^2\phi + \left(\frac{1}{r} + \frac{\partial_rn_0(r)}{n_0(r)}\right) \partial_r\phi + \frac{1}{r^2}\partial_\theta^2\phi\right] + \frac{1}{T_e(r)}(\phi-\left\langle \phi\right\rangle) = \frac{1}{n_0(r)}\int_\mathbb{R}f\,dv -1.
    \label{eq:qne}
\end{equation}
For the quasi neutrality equation, periodic boundary conditions in $(\theta,z)$, homogeneous Dirichlet boundary conditions in $r_\text{max}$ and homogeneous Neumann boundary conditions at $r_\text{min}$ are considered, i.e., we enforce no poloidal rotation of the plasma at $r_\text{min}$.

In order to solve~\eqref{eq:dk}, we first rewrite it in its conservative form by multiplying the solution with the Jacobian of the transformation from the Cartesian to the polar coordinate system in the poloidal plane. Thus we consider $g(t,r,\theta,z,v) = rf(t,r,\theta,z,v)$ and $g$ satisfies
\[
 \partial_t g - \partial_r\left(\frac{\partial_\theta \phi}{r} g\right) +\partial_\theta \left( \frac{\partial_r \phi}{r} g \right)  + \partial_z(v\,g) - \partial_v(\partial_z\phi\, g) = 0.
\]
Additionally, since $g$ can be written as $g(t) = g_{\text{eq}} + \delta g(t)$, i.e., as a sum of a steady state solution of the system and a perturbation, where $g_{\text{eq}} = rf_{\text{eq}}$, it is possible to work with $\delta g$ instead of $g$. This is done to avoid problems with boundary conditions in the radial direction, similar difficulties are observed  in~\cite{CROUSEILLES2018144,Crouseilles2014,GRANDGIRARD2006395}. Thus, we are interested to solve
\begin{equation}
 \partial_t \delta g - \partial_r\left(\frac{\partial_\theta \phi}{r} \delta g\right) +\partial_\theta \left( \frac{\partial_r \phi}{r} \delta g \right)  + \partial_z(v\, \delta g) - \partial_v(\partial_z\phi\, \delta g) - \partial_\theta \phi \partial_r\left(\frac{g_\text{eq}}{r}\right) - \partial_z \phi \partial_v g_\text{eq}= 0,
\end{equation}
where two additional source terms regarding the equilibrium function appear.
To ensure mass conservation up to machine precision, we set the velocity field component in the radial direction at $r_\text{min}$ to zero. Numerical experiments have shown that this slight modification has no significant impact on the obtained results since the velocity field is anyhow close to zero at this boundary. Due to the homogeneous Dirichlet boundary conditions imposed for the quasi neutrality equation at $r_\text{max}$, the velocity field in the radial direction is automatically zero at $r_\text{max}$. 
Since both $\delta g$ and the electric potential decay fast enough towards the radial endpoints and the velocity field is forced to be zero there, no boundary conditions for $\delta g$ in this direction have to be imposed. In the $\theta$, $z$, and $v$ dimension periodic boundary conditions are considered. Note that, when working with $\delta g$, the right-hand side of the quasi neutrality equation~\eqref{eq:qne} can be computed as 
\[
\frac{1}{n_0(r)}\int (f_\text{eq}+\delta f)\,dv -1 = \frac{1}{n_0(r)}\int \delta f \,dv = \frac{1}{r\cdot n_0(r)}\int \delta g \,dv.
\]

\section{Numerical methods and implementation}

\subsection{Time splitting \label{sec:ts}}

To solve the drift-kinetic equation numerically, splitting methods are often used, see~\cite{Crouseilles2014,GRANDGIRARD2006395} for example, to overcome the complexity of the whole problem. This allows us to treat the different parts of the equation separately. The advantage of such a procedure is that highly efficient numerical algorithms which are tailored to the simpler subproblems can be used. The drawback of such splitting methods is that, in general, an error in time is introduced. A significant amount of work can be found in the literature that is concerned with the improvement and analysis of splitting methods, see e.g.~\cite{crouseilles2015hamiltonian,einkemmer2014convergence_a,einkemmer2014convergence_b}.

The main parts (subflows) of the splitting algorithm in the drift-kinetic setting considered in this work are as follows
\begin{itemize}
 \item Solve the quasi neutrality equation and compute the derivatives of the potential $\phi$ to obtain the electric field.
 \item Perform the 1d advection in the $z$ direction with a semi-Lagrangian method, i.e., solve 
 \begin{equation}
 \label{eq:dk1dz}
\partial_t\delta g + \partial_z(v\, \delta g)=0.
 \end{equation}
 \item Perform the 1d advection in the $v$ direction with a semi-Lagrangian method, i.e., solve 
 \begin{equation}
 \label{eq:dk1dv}
 \partial_t\delta g - \partial_v(\partial_z\phi \,\delta g)=0.
 \end{equation}
 \item Perform the 2d advection in the $r$ and $\theta$ direction with a semi-Lagrangian method, i.e., solve 
 \begin{equation}
 \label{eq:dk2d}
 \partial_t \delta g - \partial_r\left(\frac{\partial_\theta \phi}{r} \delta g\right) +\partial_\theta \left( \frac{\partial_r \phi}{r} \delta g \right)=0.
 \end{equation}
 \item Treat the source terms, i.e., solve
 \[
\partial_t \delta g - \partial_\theta \phi\partial_r(g_\text{eq}/r)- \partial_z \phi \partial_v g_\text{eq}= 0.
 \]
\end{itemize}
Those five steps can be combined to obtain splitting schemes of various order. The initial condition of those five steps is the solution obtained in the previous step of the splitting procedure. The step size of each step varies according to the splitting method. 

Before proceeding let us emphasize that we treat the advection in the poloidal plane, i.e.~in $(r,\theta)$, as a 2d problem. It is possible, in principle, to split \eqref{eq:dk2d} into two 1d problems, where again a conservative SLDG scheme can be applied. This is commonly done for the Vlasov--Poisson equations, see e.g.~\cite{sonnendrucker1999semi}. 
However, since the advection speed in~\eqref{eq:dk2d} is not constant with respect to its direction, a less efficient method than will be described for~\eqref{eq:dk1dz} and~\eqref{eq:dk1dv} has to be implemented. In the case of a varying advection speed, space-dependent matrices have to be constructed, while in the constant case, just one matrix is required. Moreover, drift-kinetic problems commonly show turbulent structures in the poloidal plane, and thus the direction of advection changes appreciably from one point to another. 
Performing a further splitting then introduces a splitting error that often mandates a relatively small time step size. This is not an issue in the toroidal direction as the solution largely follows the generally well-behaved magnetic field lines. It should also be noted that, as is observed in~\cite{Cai2020comparison}, order reduction can appear in splitting advections with variable advection speed.

In this work, we will use two methods. A first order scheme, see algorithm~\ref{alg:splitting_1}, that was already considered in~\cite{Crouseilles2014}. This scheme fails to be second order because the electric field that is used to trace the characteristics is not a sufficiently accurate approximation at half the time step. This can be easily checked by a local truncation error analysis. 

We propose the second order scheme in algorithm~\ref{alg:splitting_2} that is based on a corrector-predictor strategy. 
In the Strang splitting scheme of the corrector step, the full time step is taken for the advection in the poloidal plane, which is expected to be the most costly part of the algorithm. 

\begin{algorithm}[H]
\caption{First order splitting scheme}
\label{alg:splitting_1}
\begin{enumerate}
\item solve the quasi neutrality equation to obtain~$\phi^n$ from~$\delta g^{n}$. Then, compute the velocity field and the invariants of the system,
\item treat the source term $\partial_t \delta g - \partial_\theta \phi\partial_r(g_\text{eq}/r)- \partial_z \phi \partial_v g_\text{eq}= 0$ with step size~$\Delta t/2$,
\item solve the 1d advection $\partial_t\delta g + \partial_z(v\, \delta g)=0$ with step size $\Delta t/2$, 
\item solve the 1d advection $\partial_t\delta g - \partial_v(\partial_z\phi \,\delta g)=0$ with step size~$\Delta t/2$ to get~$\delta g^\star$,
\item solve the quasi neutrality with right-hand side $\frac{1}{rn_0(r)}\int \delta g^\star \,dv$ in order to compute the potential~$\phi^\star$ to obtain the velocity field, which is used in the remaining steps,
\item solve the 2d advection $\partial_t \delta g - \partial_r\left(\frac{\partial_\theta \phi}{r} \delta g\right) +\partial_\theta \left( \frac{\partial_r \phi}{r} \delta g \right)=0$ with step size~$\Delta t$, 
\item solve the 1d advection $\partial_t\delta g - \partial_v(\partial_z\phi \,\delta g)=0$ with step size~$\Delta t/2$,
\item solve the 1d advection $\partial_t\delta g + \partial_z(v\, \delta g)=0$ with step size~$\Delta t/2$, 
\item treat the source term $\partial_t \delta g - \partial_\theta \phi\partial_r(g_\text{eq}/r)- \partial_z \phi \partial_v g_\text{eq}= 0$ with step size~$\Delta t/2$.
\end{enumerate}
\end{algorithm}
\begin{algorithm}[H]
\caption{Second order splitting scheme}
\label{alg:splitting_2}
\begin{enumerate}
\item solve the quasi neutrality equation to obtain~$\phi^n$ from~$\delta g^n$. Then, compute the velocity field and the invariants of the system,
\item [] \textbf{Predictor step, using $\phi^n$}
\item treat the source term $\partial_t \delta g - \partial_\theta \phi\partial_r(g_\text{eq}/r)- \partial_z \phi \partial_v g_\text{eq}= 0$ with step size~$\Delta t/2$ and initial condition~$\delta g^n$,
\item solve the 1d advection $\partial_t\delta g + \partial_z(v\, \delta g)=0$ with step size~$\Delta t/2$, 
\item solve the 1d advection $\partial_t\delta g - \partial_v(\partial_z\phi \,\delta g)=0$ with step size~$\Delta t/2$,
\item solve the 2d advection $\partial_t \delta g - \partial_r\left(\frac{\partial_\theta \phi}{r} \delta g\right) +\partial_\theta \left( \frac{\partial_r \phi}{r} \delta g \right)=0$ with step size~$\Delta t/2$ by using a first order characteristic tracing method to obtain~$\delta g^{n+1/2}$,
\item solve the quasi neutrality equation to obtain~$\phi^{n+1/2}$ from~$\delta g^{n+1/2}$. Then, compute the velocity field,
\item [] \textbf{Corrector step, using $\phi^{n+1/2}$}
\item treat the source term $\partial_t \delta g - \partial_\theta \phi\partial_r(g_\text{eq}/r)- \partial_z \phi \partial_v g_\text{eq}= 0$ with step size~$\Delta t/2$ with initial condition~$\delta g^n$,
\item solve the 1d advection $\partial_t\delta g + \partial_z(v\, \delta g)=0$ with step size~$\Delta t/2$, 
\item solve the 1d advection $\partial_t\delta g - \partial_v(\partial_z\phi \,\delta g)=0$ with step size~$\Delta t/2$,
\item solve the 2d advection $\partial_t \delta g - \partial_r\left(\frac{\partial_\theta \phi}{r} \delta g\right) +\partial_\theta \left( \frac{\partial_r \phi}{r} \delta g \right)=0$ with step size~$\Delta t$ by using a second order characteristic tracing method, 
\item solve the 1d advection $\partial_t\delta g - \partial_v(\partial_z\phi \,\delta g)=0$ with step size~$\Delta t/2$,
\item solve the 1d advection $\partial_t\delta g + \partial_z(v\, \delta g)=0$ with step size~$\Delta t/2$, 
\item treat the source term $\partial_t \delta g - \partial_\theta \phi\partial_r(g_\text{eq}/r)- \partial_z \phi \partial_v g_\text{eq}= 0$ with step size~$\Delta t/2$, to obtain~$\delta g^{n+1}$, which is a second order solution in time.
\end{enumerate}
\end{algorithm}

\subsection{Approximation space}
To approximate functions $u(t^n,x)\approx u^n(x)$ in space, a discontinuous Galerkin representation is considered. In 1d, Lagrangian functions that interpolate at Gauss--Legendre points are used as (orthogonal) basis functions. A function $u$ (in 1d) is then approximated as follows
\[
 u^n(x) \approx \sum_{i=1}^N\sum_{j=1}^{k+1} u^n(x_{i,j})\varphi_{i,j}(x),
\]
where $k$ is the degree of the polynomial approximation, $x_{i,j}$ is the $j^{\text{th}}$ Gauss--Legendre point scaled to cell~$i$ and where~$\varphi_{ij}$ is the Lagrange basis function corresponding to the point~$x_{i,j}$. Note that $\varphi_{ij}$ are non-zero only in cell~$i$. Therefore the degrees of freedom are the values of the function at the Gauss--Legendre points.
For higher dimensional problems tensor products of the 1d basis functions are used. Therefore, for example in 2d, the domain is divided into rectangular cells and the basis functions in each cell~$(i,j)$ are
\[
 \varphi_{ij,m}(x,y) =  \varphi_{i,p}(x)\varphi_{j,q}(y),
\]
where $m = m(p,q)$ is a linearized index of $(p,q)$. This can be extended to arbitrary dimensions. Thus, for the problem considered here in the 4d setting, the domain is divided into four dimensional cells. In each such cell the density function is approximated by a polynomial.

\subsection{Semi-Lagrangian discontinuous Galerkin in 1d with constant coefficients}
\begin{figure}[H]
\begin{center}
 \includegraphics[width=0.8\textwidth]{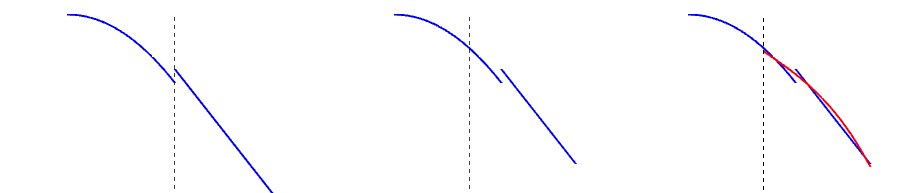}
 \caption{Illustration of the one dimensional semi-Lagrangian discontinuous Galerkin method. When the advective step is completed (figure in the middle), the solution is projected back to the approximation space (red curve in the right figure).}
 \label{fig:dg1d}
 \end{center}
\end{figure}

To solve the two 1d advection problems~\eqref{eq:dk1dz} and~\eqref{eq:dk1dv}, a semi-Lagrangian discontinuous Galerkin method is used. Since the advection speed for the two 1d advections in consideration does not depend on the direction where the advection takes place, an extremely efficient algorithm can be derived. For each degree of freedom in the 3d subspace where the advection is not performed, the following problem has to be solved,
\begin{equation}
\partial_t u + \partial_x(a u) = 0.
\label{eq:sldg1d_conservative}
\end{equation}
Note that with an advection velocity that does not depend on $(t,x)$ we can pass from the conservative to the advective form of the transport equation without any additional considerations. The exact solution of this problem is known, namely $u(t,x) = u_0(x-at)$, which can be derived with the method of characteristics. The numerical scheme uses the same idea, namely following the characteristics backward in time. However, since the endpoint of a characteristic curve does not necessarily coincide with a grid point, an interpolation technique has to be applied. In our case, since the approximation space consists of piece-wise polynomial functions which are discontinuous at the cell interface, the shifted function is no longer a continuous polynomial in the cells. Therefore, an $L^2$ projection is applied in order to remain in the approximation space. The idea of this method is shown in figure~\ref{fig:dg1d}.
To derive the variational form, the following theorem can be used,
\begin{equation}
 \partial_t \left(\int_{A_i(t)}u(t,x)\varphi(t,x)dx\right) = 0,
 \label{eq:sldg_theorem}
\end{equation}
where $A_i(t)$ is the moving Lagrangian cell in time and where $\varphi$ solves the adjoint problem, 
\begin{equation}
\partial_t \varphi + a\partial_x \varphi = 0, \quad \varphi(t^{n+1}) \text{  is given.}
\label{eq:adjoint_problem}
\end{equation}
For a proof we refer to~\cite{Guo2014}. Then evaluating~\eqref{eq:sldg_theorem} at time $t^n$ and $t^{n+1}$ gives 
\begin{equation}
 \int_{A_i}u^{n+1}\varphi(t^{n+1},x)\,dx = \int_{A_i(t^n)}u^{n}\varphi(t^n,x)\,dx.
\end{equation}
Since in the one dimensional case the advection speed is constant, $A_i(t^n)$ and $\varphi(t^n,x)$ can be determined exactly and are just shifted versions of their initial conditions. As a consequence, $A_i(t^n)$ overlaps always with at most two underlying Eulerian cells and in order to find those cells, following the endpoints of the interval $A_i$ along the characteristic is sufficient. This results in the following algorithm
\begin{equation}
 u^{n+1}_{i,\cdot} = A(a)u^{n}_{i^*,\cdot} + B(a)u^{n}_{i^*+1,\cdot},
 \label{eq:sldg1d_implement}
\end{equation}
where $A(a)$ and $B(a)$ are small matrices of size $(k+1)\times (k+1)$ and $k$ is the degree of the polynomial used. The index $i^*$ gives the cell where the left endpoint of the interval $A(t^n)$ lies after following the characteristics. These matrices do depend only on the advection speed and can thus be precomputed.
See~\cite{crouseilles2011discontinuous} for more details about the derivation of the method and the form of the matrix.
The method is of order $k+1$ in space and is mass and momentum conservative by construction. Moreover, since the advection speed is constant and the characteristics can be computed analytically and never cross, this scheme is unconditionally stable. Furthermore, comparable and in some cases even less numerical diffusion is introduced than by cubic splines, see~\cite{Einkemmer20194d}.

This method can be implemented extremely efficiently. In each step, all degrees of freedom have to be read and written once, and for each cell just the sum of two small matrix-vector products has to be performed. Moreover, since only two adjacent cells are required to compute the values of a cell at the next time step, this scheme can be efficiently and easily parallelized and is thus interesting from a high performance computing point of view. More details about the efficiency of the implementation on multi-core CPU and GPU based systems can be found in~\cite{Einkemmer20194d,Einkemmer2020GPUs,Einkemmer2022}. 

\subsection{Semi-Lagrangian discontinuous Galerkin in 2d}
To solve the 2d advection~\eqref{eq:dk2d}, again a semi-Lagrangian discontinuous Galerkin method is used. This recently developed method in~\cite{Cai2017} conserves mass up to machine precision, which is not the case for the standard backward semi-Lagrangian method used in~\cite{GRANDGIRARD2006395}. This is an important property, especially for long time simulations, as it helps to obtain physically relevant results and to improve the quality of the solution. We recall in the following the main ideas of the method developed in~\cite{Cai2017,Lauritzen2010}, which is used to solve~\eqref{eq:dk2d}.

\begin{figure}[H]
\begin{center}
 \includegraphics[width=0.9\textwidth]{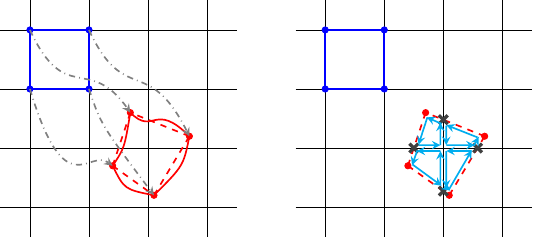}
 \caption{Illustration of the SLDG 2d algorithm. Left picture: Tracing the characteristics of the corner points of $A_{ij}$  (gray arrow), then, $A^*_{ij}$ is created by connecting the corner points with straight lines (red dashed lines), which is a second order approximation of $A_{ij}(t)$ (red lines). Right picture: The intersection points with the Eulerian grid cells have to be found (black crosses), and finally, the coefficients~\eqref{eq:sldg2d_coeff} are computed by integrating over $A^*_{ij,rs}$ and using Green's theorem (cyan arrows).}
 \label{fig:sldg2d}
\end{center}
\end{figure}

We are interested to solve problems of the following form,
\begin{equation}
 \partial_t u + \partial_x(a(t,x,y)u) + \partial_y(b(t,x,y)u) = 0.
 \label{eq:sldg2d_conservative}
\end{equation}
Similarly as in 1d, the 2d extensions of equations~\eqref{eq:sldg_theorem} and~\eqref{eq:adjoint_problem} are used to derive the variational form
\begin{equation}
\int_{A_{ij}}u(t^{n+1},x,y)\varphi(t^{n+1},x,y)\,dxdy = \int_{A_{ij}(t^n)}u(t^{n},x,y)\varphi(t^{n},x,y)\,dxdy,
\label{eq:sldg2d_variationalform}
\end{equation}
where the velocity field in our case depends on $t$, $x$, and $y$. This implies that additional difficulties arise since we can not compute, in general, exactly the upstream cell $A_{ij}(t^n)$, see the left picture in figure~\ref{fig:sldg2d}, and the analytical solution of the adjoint problem, $\varphi(t^n,x,y)$, is not known in general.

To find the upstream cell by following the corner points of an Eulerian grid cell along the characteristics back in time and connecting them with straight lines introduces an error of order two in space.  There exist several strategies to obtain a higher order approximation of the upstream cell, for example, using higher order polynomial approximations of the faces, see~\cite{Cai2017}, or using so called Eulerian-Lagrangian schemes, which introduce additional terms to correct the error which is coming from the boundary, see~\cite{Cai2021EL}. Let us note that the error in space for semi-Lagrangian schemes is related to the error in time when tracing the characteristics. Since we are working with an at most second order in time splitting method, we expect to get at most an error of order two in space.
Therefore, we use quadrilaterals in this article to approximate the upstream cells and polynomial approximations of order two. We denote the approximated upstream cell by $A^*_{ij}$.
Let us mention that if the characteristics can be traced exactly, in general large time steps can be taken. However, if this is not the case, which happens for example in problems where the velocity field is not analytically available, approximations of the characteristics have to be computed. This leads to time step restrictions due to a loss of accuracy and even stability problems might appear if the approximated characteristics cross each other. By using higher order methods to trace the characteristics, as is done in~\cite{Cai2021}, large time steps can be used. 

To solve the adjoint problem, some points of the cell are chosen and the characteristics are then followed backward in time starting from those points. Since the solution stays constant on these points, $\varphi(t^n,x,y)$ is then approximated by a polynomial $\varphi^*(x,y)$ with an appropriate strategy. 

\subsubsection*{Lagrange vs Legendre basis functions}

For discontinuous Galerkin approximations in 2d, if the domain is divided into quadrilaterals, polynomials in $\mathbb{P}^k$ are enough to get a method of order $k+1$ (in contrast to finite element approximations, see~\cite{dgbook}). Therefore, as basis often Legendre polynomials are used due to their favorable orthogonality properties. Also in~\cite{Cai2017}, the algorithm is implemented in this way. 

In this work, we use tensor products of Lagrange functions, as explained previously, as a basis for the two dimensional subspace. This is done to have a consistent representation along all four dimensions. In the literature, this polynomial space is often referred as~$\mathbb{Q}^k$, which is larger than~$\mathbb{P}^k$, since it has in two dimensions~$(k+1)^2$ degrees of freedom in each cell in contrast to~$(k+1)(k+2)/2$. This implies that more memory is required and this is consequently an issue for high dimensional problems. However, since more degrees of freedom are used, in general the resulting error is also lower, as has been observed in~\cite{Cai2020comparison}. In this article, we use $k=1$.

For different basis functions in the 2d SLDG method, different strategies to find approximations of the adjoint problem $\varphi(t^n,x,y)$ have to be applied. For Legendre basis functions (as considered in \cite{Cai2017}), the four corner points of the cell are traced along the characteristics backward in time. Since the solution of the adjoint problem (2d extension of~\eqref{eq:adjoint_problem}) stays constant at the characteristics, an approximating polynomial of the solution can be constructed. As explained above, a second order method results in three degrees of freedom in each cell. Therefore, more backward traced points (four) than degrees of freedom (three) are given and consequently a least square strategy is applied to minimize the error for the constructed approximating polynomial. 

By applying a similar strategy that uses the corner points when Lagrange polynomials are used, we observe numerical instabilities when the upstream cell is no longer a rectangle, i.e., we observe wrong and fast increasing slopes of the polynomial in each cell. Thus, instead of using the corner points of the upstream cell, we use the four points $(\xi^n_m,\eta^n_m)_{m=1,\ldots,4}$ that are obtained by tracing the characteristics back in time starting from the Gauss--Legendre nodes $(\xi^{n+1}_m,\eta^{n+1}_m)_{m=1,\ldots,4}$ of a cell. Then, we solve a small $4\times 4$ Vandermonde system, i.e., $\varphi^*(\xi^n_m,\eta^n_m) = \varphi(t^{n+1},\xi^{n+1}_m,\eta^{n+1}_m)$ for $m=1,\ldots,4$, to obtain the second order polynomial approximation $\varphi^*(x,y)=\sum_{i,j=0}^{k} c_{ij}x^iy^j$ of~$\varphi(t^n,x,y)$. This strategy is more natural as the Lagrange basis functions are defined on the Gauss--Legendre nodes and we observe numerically robust simulations.

\subsubsection*{Characteristics tracing}

In order to trace the characteristics of~\eqref{eq:sldg2d_conservative} backward in time, we consider the following first order method
\begin{equation}
\label{eq:characteristics-1}
\left\{
\begin{aligned}
x^n &= x^{n+1} - \Delta t\cdot a(x^{n+1},y^{n+1},t^n) \\
y^n &= y^{n+1} - \Delta t\cdot b(x^{n+1},y^{n+1},t^n)
\end{aligned}
\right.
\end{equation}
and a second order method,
\begin{equation}
\label{eq:characteristics-2}
\left\{
\begin{aligned}
x^n &= x^{n+1} - \Delta t\cdot a(x^{n+1/2},y^{n+1/2},t^{n+1/2}) \\
y^n &= y^{n+1} - \Delta t\cdot b(x^{n+1/2},y^{n+1/2},t^{n+1/2})
\end{aligned}
\right.
\end{equation}
respectively, where the starting points $x^{n+1}$ and $y^{n+1}$ are either the cell corners or the Gauss--Legendre points inside the cells.
The first order scheme~\eqref{eq:characteristics-1} can be easily implemented since the velocity field at time $t^n$ is known. 
For the second order method~\eqref{eq:characteristics-2} additional considerations have to be taken.
First, half a step size has to be computed in advance in order to get the velocity field at time $t^{n+1/2}$ by solving the quasi neutrality equation. This is done in the predictor step of the time splitting method and thus no additional cost is incurred. Then we can use the first order scheme~\eqref{eq:characteristics-1} to compute $x^{n+1/2}$ and $y^{n+1/2}$. Finally, to obtain the velocity field at these points, we have to perform an interpolation since the velocity field is only known at the grid points. The resulting order can be checked with a local truncation error analysis. Although there exist higher order methods in the literature to trace the characteristics, see~\cite{Cai2021,Cai2019} for example, we restrict ourselves to the second order scheme since the time splitting introduces an error of order two in any case.

\subsubsection*{Description of the algorithm}

In order to implement~\eqref{eq:sldg2d_variationalform}, the following steps have to be taken. 
\begin{itemize}
 \item Trace the corner points of the Eulerian grid cells $A_{ij}$ along the characteristics back in time to find a second order approximation~$A^*_{ij}$ of the upstream cell~$A_{ij}(t^n)$, see the left picture of figure~\ref{fig:sldg2d}.
 \item Determine a polynomial approximation $\varphi_{ij,m}^*$ of the solution of the adjoint problem, where the initial condition are the basis functions, by tracing the Gauss-Legendre points along the characteristics back in time and solve the corresponding small Vandermonde system.
 \item Approximate~\eqref{eq:sldg2d_variationalform} by 
 \begin{equation}
 \label{eq:sldg2d_splitintegral}
 \begin{aligned}
  \int_{A_{ij}}u^{n+1}(x,y)\varphi_{ij,m}(t^{n+1},x,y)\,dxdy &\approx 
\int_{A^*_{ij}}u^{n}(x,y) \varphi_{ij,m}^*(x,y)\,dxdy \\
&= \sum_{rs} \int_{A^*_{ij,rs}}u^n(x,y) \varphi_{ij,m}^*(x,y)\,dxdy .
\end{aligned}
\end{equation}
 where $A^*_{ij,rs}$ is the intersection of the upstream cell with the underlying Eulerian grid cells $A_{rs}$. In order to determine them, the intersection points between $A^*_{ij}$ and the Eulerian cells have to be found, see the black crosses in figure~\ref{fig:sldg2d}.
 \item Finalize by writing~\eqref{eq:sldg2d_splitintegral} as
 \begin{equation}
 u^{n+1}_{ij,m}w_m = \sum_{rs}\sum_k u^{n}_{rs,k}C^{ij,m}_{rs,k},
 \label{eq:sldg2d_implement}
 \end{equation}
 where
 \begin{equation}
  C^{ij,m}_{rs,k} = \int_{A^*_{ij,rs}} \varphi_{rs,k}(x,y)\varphi_{ij,m}^*(x,y)\,dxdy.
  \label{eq:sldg2d_coeff}
 \end{equation}
 These coefficients are then computed by using Green's theorem,
 \[
  \int_{A^*_{ij,rs}} \varphi_{rs,k}(x,y)\varphi_{ij,m}^*(x,y)\,dxdy = \oint_{\partial A^*_{ij,rs}} [P(x,y)dx + Q(x,y)dy],
  \]
 where
 \[
 -\partial_yP(x,y)+\partial_xQ(x,y) = \varphi_{rs,k}(x,y)\varphi_{ij,m}^*(x,y) = \sum_{p,q=0}^{2k}c_{(p,q)}x^py^q
 \]
 In order to do so, the surrounding faces of $A^*_{ij,rs}$ have to be determined, see the cyan arrows in figure~\ref{fig:sldg2d}. More precisely, 
 \begin{align*}
  \int_{A^*_{ij,rs}} \varphi_{rs,k}(x,y)\varphi_{ij,m}^*(x,y)\,dxdy &= \sum_{p,q=0}^{2k}c_{(p,q)}\int_{A^*_{ij,rs}} x^py^q\,dxdy \\
  &= \sum_{p,q=0}^{2k}c_{(p,q)} \sum_l \int_{\partial A^*_{ij,rs,l}} [P^{(p,q)}(x,y)dx + Q^{(p,q)}(x,y)dy]
 \end{align*}
 where $\cup_l \partial A^*_{ij,rs,l} = \partial A^*_{ij,rs}$ and the functions $P^{(p,q)}$ and $Q^{(p,q)}$ are chosen for $p+q\leq2$ as in~\cite{Lauritzen2010} and the higher degree polynomials are chosen as follows
 \begin{align*}
 P^{(2,1)} = 0, \quad
 Q^{(2,1)} = \frac{x^3y}{3}, \quad
 P^{(1,2)} = -\frac{xy^3}{3}, Q^{(1,2)} = 0, \quad
 P^{(2,2)} = 0, Q^{(2,2)} = \frac{x^3y^2}{3}. 
 \end{align*}
 Let us mention that the choice of $P^{(p,q)}$ and $Q^{(p,q)}$ is not unique, but its integral is.
\end{itemize}
 Although the numerical implementation in~\eqref{eq:sldg2d_implement} looks similar to~\eqref{eq:sldg1d_implement} in the one dimensional case, i.e., as a sum of matrix vector products, it is much more challenging to get good performance from an implementation on GPUs. In contrast to the 1d case, several differences appear. First, the velocity field varies in space, which implies that for each cell different matrices have to be computed. Additionally, since the computation of these quantities does depend on the geometry determined by following the characteristics, a lot of control flow is required, which is a scarce resource on GPUs. Second, since the advection speed is not constant, it is not clear a priori how many grid cells are intersected by the upstream cell. This is especially a problem on GPUs with their small caches since the loaded data can often not be reused. Moreover, as explained previously, the amount of data to read varies from cell to cell and the memory access pattern can not be predicted. Third, since this problem is two dimensional, we have to deal with a strided memory access.

However, for realistic time step sizes the area of the upstream cell does not vary too much. Then, a reasonably small range of $rs$ in~\eqref{eq:sldg2d_implement} can be chosen. Thus, as in the 1d case, it follows that just a few cells are required to compute the values at the next time step for a given cell. Therefore, this method is local which is advantageous in a high performance computing setting (especially if one wishes to scale such methods to large systems).

\subsection{Source term}
As explained previously, the advective splitting substeps are mass conservative. In order to obtain a global mass conservative method, also the way how the source term is treated has to be mass conservative. If we integrate the term $\partial_z \phi \partial_v g_\text{eq}$ over the full domain, we obtain
\begin{align*}
\text{mass} &= \int_{r_\text{min}}^{r_\text{max}}\int_0^{2\pi}\int_0^L\int_{-v_\text{max}}^{v_\text{max}}  \partial_z \phi(t,r,\theta,z) \partial_v g_\text{eq}(r,v)\,drd\theta dzdv \\
&= \int_{r_\text{min}}^{r_\text{max}}\int_0^{2\pi}\int_{-v_\text{max}}^{v_\text{max}}  \partial_v g_\text{eq}(r,v)\left(\int_0^L \partial_z \phi(t,r,\theta,z)\,dz\right) \,drd\theta dv \\
&= \int_{r_\text{min}}^{r_\text{max}}\int_0^{2\pi}\int_{-v_\text{max}}^{v_\text{max}}  \partial_v g_\text{eq}(r,v) \left(\phi(t,r,\theta,L)-\phi(t,r,\theta,0)\right) \,drd\theta dv = 0
\end{align*}
due to the periodicity of $\phi$ in the $z$ direction. A similar computation holds for the other term, $\partial_\theta \phi \partial_r\left(\frac{g_\text{eq}}{r}\right)$. Therefore, if the derivative is discretized in such a way as to preserve this property, we obtain a mass conservative scheme. This is, e.g., the case if we use centered finite differences to compute the derivatives. Then, the source term can be simply treated as
\[
 \delta g^{\star,n+1} = \delta g^{\star,n} + \Delta t\left( \partial_z \phi \partial_v g_\text{eq} + \partial_\theta \phi \partial_r\left(\frac{g_\text{eq}}{r}\right)\right)
\]
in the splitting step and does not introduce an error in mass.

\subsection{Quasi neutrality equation}
For the quasi neutrality equation, periodic boundary conditions are considered in the $\theta$ and $z$ direction, homogeneous Dirichlet boundary conditions in the radial direction at $r_\text{max}$ and homogeneous Neumann boundary conditions at $r_\text{min}$ are considered. Therefore, we can use FFT in the $(\theta,z)$ plane. In addition, we use finite differences in the $r$ direction. Thus, for each degree of freedom in the $\theta$ and $z$ direction, a tridiagonal linear system has to be solved. See~\cite{GRANDGIRARD2006395} for more details. Since we are working with a non equidistant nodal basis, in order to use the FFT, we have to evaluate first the right-hand side of the equation, i.e., $\frac{1}{rn_0(r)}\int \delta g^n \, dv$, at an equidistant grid. In the $z$ direction, this is done inside the cell, while in the $r$-$\theta$ plane it is done at the cell corners, where the average is taken since the function is discontinuous here. To compute the velocity field, we use again second order finite differences. In order to obtain the values of the transport field inside the cells, bilinear interpolation is applied. 

\subsection{Implementation details}

Since the 2d advection is the most expensive part of solving the 4d problem, memory is organized to favor memory access in that step. Thus, the fastest varying indices in memory correspond to $r$ and $\theta$. In addition, all degrees of freedom in a cell are stored consecutively in memory. The next direction in memory is the $v$ dimension and finally the $z$ dimension is considered. 
This choice is made because the velocity field also depends on $z$ and thus the coefficients~\eqref{eq:sldg2d_coeff} need to be recomputed for each $z$.
This leads to an implementation as in algorithm~\ref{alg:sldg2d_algo_general} which allows us to reuse those coefficients 
for each $v$.
\begin{algorithm}[H]
\begin{program}
\FOR z \DO
    |compute all | C^{ij,m}_{rs,k}, | see | \eqref{eq:sldg2d_coeff}
    \FOR v \DO
       |perform the advection in 2d, see | \eqref{eq:sldg2d_implement} 

\end{program}
\caption{2d advection in the drift-kinetic setting}
\label{alg:sldg2d_algo_general}
\end{algorithm}
We have implemented this both on the CPU and on the GPU. While for the CPU the implementation is straightforward by parallelizing first the computation of~\eqref{eq:sldg2d_coeff} and then the loop over the degrees of freedom in the $v$ direction using OpenMP, the implementation on the GPU requires more care in order to obtain good performance. 

On the GPU algorithm~\ref{alg:sldg2d_algo_general} is divided into two steps. The first part consists of computing for each degree of freedom in the $z$ direction the coefficients $C^{ij,m}_{rs,k}$. The second part consists of one kernel call which implements~\eqref{eq:sldg2d_implement} for the full domain. Here, threads within a block are reading the coefficients~\eqref{eq:sldg2d_coeff} of one cell in the $r$-$\theta$ domain to the shared memory. The number of threads per block is given by the maximum theoretical coefficients that have to be loaded. In our case, since the approximation space with order two has four elements, this implies that for each possible intersection of the upstream cell with the underlying Eulerian grid cell a matrix of $4\times4$ has to be read. We assume, that an upstream cell intersects at most a block of $3\times3$ grid cells, which implies that a block consists of at most 144 threads. When the coefficients are loaded, each thread is associated to one specific $v$-value to perform the advection. In other words, fine-grained thread parallelization is performed in the velocity dimension to optimize the use of the shared memory, while coarse block parallelization is performed at the cells in the $r$-$\theta$ plane and over the $z$ dimension. The 2d SLDG code is part of \texttt{sldg} which is available at \texttt{https://bitbucket.org/leinkemmer/sldg}.

\section{Numerical simulation}

\subsection{Validation of the 2d SLDG algorithm with Lagrange polynomials}

To demonstrate that the method produces the expected results also in the case of Lagrange basis functions, where we apply a different strategy in approximating the solution of the adjoint problem as described in the previous section, we consider the swirling deformation test case. This test is commonly used to demonstrate the correctness of newly developed algorithms as the advection field depends on space and time and the exact solution at the final time~$t=T$ is the same as at time~$t=0$. The problem is stated as follows 
\[
 \partial_t u - \partial_x\left(\cos^2(x/2)\sin(y)g(t)u\right) + \partial_y(\sin(x)\cos^2(y/2)g(t)u)=0,
\]
where $g(t) = \cos(\tfrac{\pi t}{T})\pi$. The initial condition is a $C^5$ cosine bell. More details about this problem can be found in~\cite{Cai2017}. The results are presented in table~\ref{tab:swirling}, where we observe the expected second order convergence rate for k=1.

\begin{table}[H]
\begin{center}
 \begin{tabular}{c|r|c|r|c}
 mesh size & $L^2$ error & $L^2$ order & $L^{\infty}$ error & $L^\infty$ order  \\ \hline
 $20\times 20$   & 5.20e-2 &      & 1.57e-1 & \\
 $40\times 40$   & 9.94e-3 & 2.39 & 4.49e-2 & 1.80 \\
 $80\times 80$   & 1.85e-3 & 2.42 & 8.66e-3 & 2.37 \\
 $160\times 160$ & 3.78e-4 & 2.29 & 1.57e-3 & 2.46 
 \end{tabular}
\caption{Swirling deformation test case using a time step of $\Delta t = 0.5\Delta x$ and a final time $T=1.5$. The error in mass in all simulations is close to machine precision.}
\label{tab:swirling}
\end{center}
\end{table}

\subsection{Drift-kinetic ion temperature gradient instability}

In this section we consider the simulation of an ion temperature gradient instability using the full four dimensional drift-kinetic problem specified in equations \eqref{eq:dk} and \eqref{eq:qne}. The initial condition and parameters are given as follows
\begin{align*}
f(t=0,r,\theta,z,v) &= f_{\text{eq}}(r,v)\left[1+\epsilon  \exp\left(-\frac{(r-r_p)^2}{\delta_r}\right)\cos\left(\frac{2\pi n}{L}z+m\theta\right)\right] \\
f_{\text{eq}}(r,v) &=\frac{n_0(r)\exp\left(-\frac{v^2}{2T_i(r)}\right)}{(2\pi T_i(r))^{1/2}}\\
\mathcal{P}(r) &= C_\mathcal{P}\exp\left(-\kappa_\mathcal{P}\delta r_\mathcal{P} \tanh\left(\frac{r-r_p}{\delta r_\mathcal{P}},  \right) \right), \quad \mathcal{P} \in \{T_i, T_e, n_0\} \\
C_{T_i}&=C_{T_e}=1, C_{n_0} = \frac{r_{\text{max}}-r_{\text{min}}}{\int^{r_{\text{max}}}_{r_{\text{min}}}\exp(-\kappa_{n_0}\delta r_{n_0} \tanh(\frac{r-r_p}{\delta r_{n_0}}))\,dr } \\
r_{\text{min}} &=0.1, r_{\text{max}}=14.5, \kappa_{n_0}=0.055, \kappa_{T_i}=\kappa_{T_e}=0.27586, \\
\delta r_{T_i} &= \delta r_{T_e} = \frac{\delta r_{n_0}}{2} = 1.45, \epsilon = 10^{-6}, n=1, m=5,\\
L&=1506.759067, r_p = \frac{r_{\text{max}}+r_{\text{min}}}{2}, \delta r = \frac{4\delta r_{n_0}}{\delta r_{T_i}}.
\end{align*}
In addition to the distribution function $f$ and the potential $\phi$ we also consider the time evolution of the electric energy
\[
 \sqrt{\int \phi^2(t,r_p,\theta,z)\,d\theta dz},
\]
the total mass
\[
 \int f(t,r,\theta,z,v)r\,drd\theta dzdv,
\]
the $L^2$ norm
\[
 \sqrt{\int f^2(t,r,\theta,z,v)r\,drd\theta dzdv},
\]
and the total energy
\[
 \int \left(v^2/2 + \phi(t,r,\theta,z)\right) f(t,r,\theta,z,v)r\,drd\theta dzdv.
\]
The total mass, the $L^2$ norm and the total energy are conserved quantities of the system. 
Grids of $64\times64\times64\times128$ and $82\times82\times82\times164$ cells and fixed time step sizes of 2,4 and 8 are used.
Let us emphasize that there is no CFL condition; that is, even larger time steps can be taken while maintaining stability. Therefore, the time step size is not restricted by the stability of the scheme, but by the accuracy of the obtained results.

\begin{figure}[H]
\begin{subfigure}{0.5\textwidth}
\begin{center}
\includegraphics[width=1.0\textwidth]{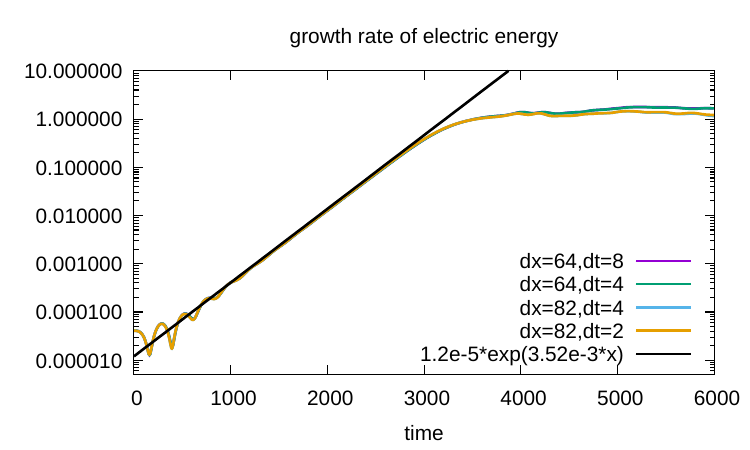}
\caption{Time evolution of the electric energy.}
\label{fig:electricenergy}
\end{center}
\end{subfigure}
\begin{subfigure}{0.5\textwidth}
\begin{center}
\includegraphics[width=1.0\textwidth]{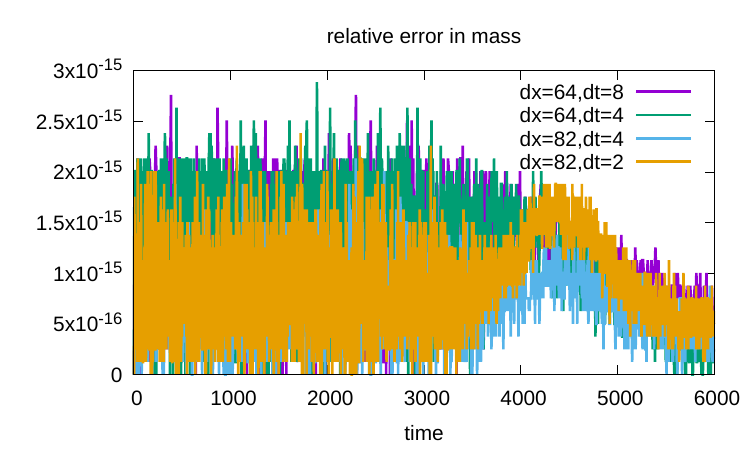}
\caption{Relative error in mass.}
\label{fig:dkmass}
\end{center}
\end{subfigure}
\vskip\baselineskip
\begin{subfigure}{0.5\textwidth}
\centering
\includegraphics[width=1.0\textwidth]{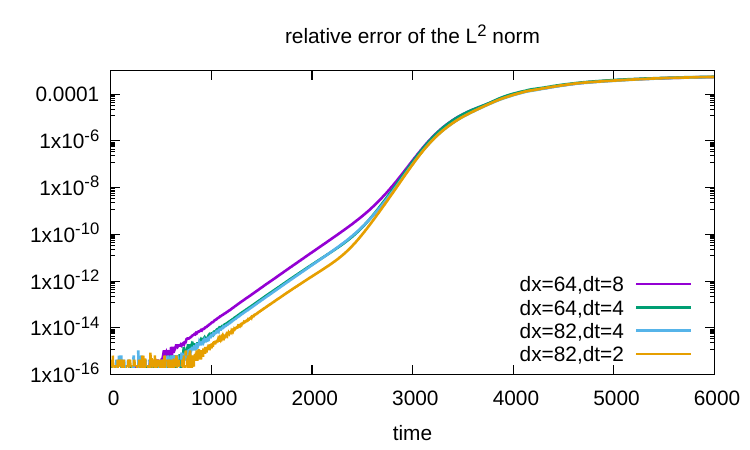}
\caption{Relative error of the $L^2$ norm.}
\label{fig:dkl2}
\end{subfigure}
\begin{subfigure}{0.5\textwidth}
\centering
\includegraphics[width=1.0\textwidth]{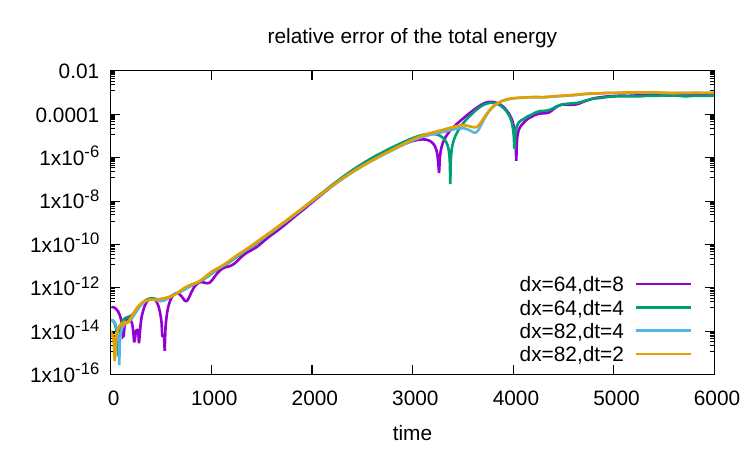}
\caption{Relative error of the total energy. }
\label{fig:dktotenergy}
\end{subfigure}
\caption{Time evolution of the electric energy as well as error in mass, $L^2$ norm and total energy is shown. A time step size of 2,4 and 8 a grid of $64\times64\times64\times128$ (dx=64) and a grid of $82\times82\times82\times164$ (dx=82) cells with order 2 is used.}
\end{figure}

Due to the gradient in temperature given by $T_i$ a physical instability develops that results in an exponential increase in the electric energy. This increase is maintained until saturation is caused by a strongly nonlinear dynamics. The initial increase of the electric energy can be obtained analytically by solving the linearized drift-kinetic model, see~\cite{Coulette2013}. In figure~\ref{fig:electricenergy}, it can be observed that the simulated growth rate matches the analytical rate very well. 

The numerical methods considered are mass conservative, but do not conserve the $L^2$ norm and the total energy. In figure~\ref{fig:dkmass} we observe that mass is indeed conserved up to machine precision. From figures~\ref{fig:dkl2} and~\ref{fig:dktotenergy} we see that the $L^2$ norm and the total energy are approximately conserved up to an error of $10^{-4}$.   

\begin{figure}[H]
\centering
 \includegraphics[width=1.0\textwidth]{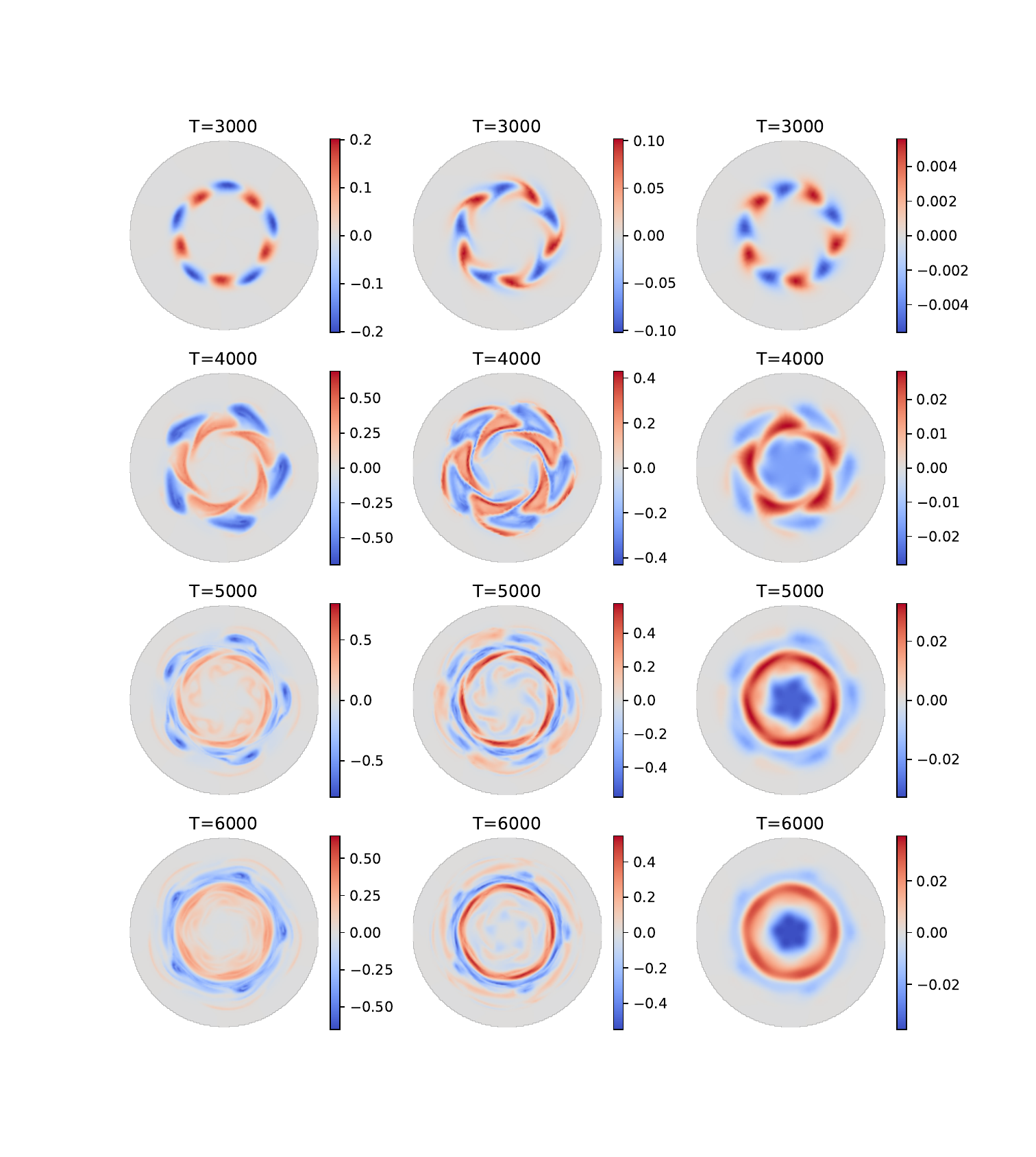}
 \caption{Plots of slices of the density $\delta g$ at $v=0$, $z=0$ (first column), the charge density at $z=0$ (second column) and electric potential at $z=0$ (third column) at different times using a grid of $64\times64\times64\times128$ cells and a time step of 4.}
 \label{fig:heatplots}
\end{figure}

Finally, in figure~\ref{fig:heatplots} we present slices of the density function $\delta g$ at $(r,t,v=0,z=0)$, the charge density and the electric potential at $(r,t,z=0)$ at time $T=3000, 4000, 5000, 6000$. That is, from the late stages of the linear instability to the solution in the fully nonlinear regime. The results match well with what has been reported in literature~\cite{GRANDGIRARD2006395, crouseilles2020exponential}.

\subsubsection*{Convergence order in time}

\begin{figure}[H]
\begin{center}
\includegraphics[width=0.8\textwidth]{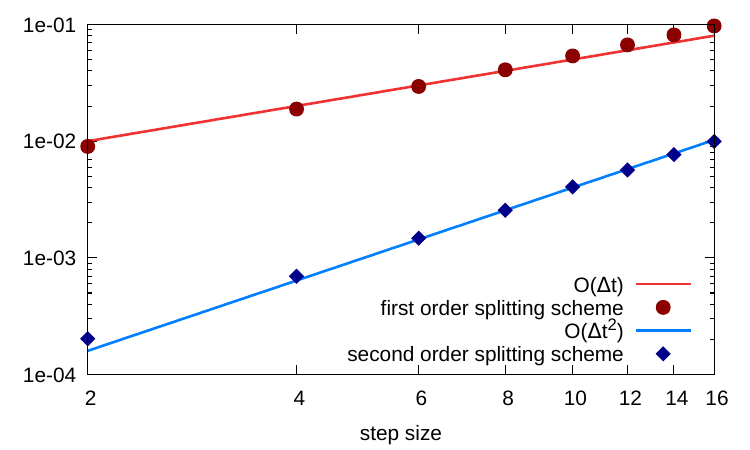}
 \caption{Convergence orders for the method from \cite{Crouseilles2014} and the proposed second order splitting scheme. The relative error in the maximum norm at the final time $T=500$ is shown.}
 \label{fig:order_in_time}
\end{center}
\end{figure}

To check the achieved order in time, we integrate the problem up to time $T=500$ and use a fixed discretization of 64 cells in the spatial dimensions and 128 cells in the velocity dimension. As reference solution, we take the result of the second order time splitting method using $2000$ time steps, which corresponds to a time step size of $0.25$. The relative error in maximum norm as a function of the step size is shown in figure~\ref{fig:order_in_time}. As expected, based on the discussion in section \ref{sec:ts} the method from \cite{Crouseilles2014} is only first order accurate, whereas the method proposed in this work is second order accurate. 
 
\section{Multi-core CPU and GPU performance Results}

In this section, we will study the performance of the implementation on multi-core CPU and GPU based systems.

Let us first discuss if the problem considered is compute bound or memory bound. While compute bound problems often appear in machine learning, i.e., performance is dictated by how many arithmetic operations can be performed, most of the scientific codes (stencil codes, sparse matrix operations, etc.) are memory bound, i.e., the limiting factor to achieve faster execution times is the time required to read from and write to memory. As is pointed out in~\cite{Einkemmer2020GPUs,Einkemmer2022}, performing the 1d advections is a memory bound problem, since comparably few floating point operations have to be performed compared to the amount of memory access.

Regarding the 2d SLDG implementation, the final step requires in each cell at most nine matrix vector products of size $4\times4$, which results in $4[9(2\cdot4-1)+8]=284$ floating point operations per cell. Since the entire domain has to be read and written once, an arithmetic intensity, when using double precision, of 
\[
 \frac{\text{dof}\cdot [9(2\cdot4-1)+8]}{2\cdot \text{dof}\cdot 8} = 4.44
\]
is obtained in the worst case. Thus also the 2d SLDG algorithm is a memory bound problem on the GPUs considered in this work, see table~\ref{tab:GPUs}. 
\begin{table}[H]
\centering
 \begin{tabular}{l|c|r|c}
 GPU     & memory & bandwidth & flops/byte (double) \\ \hline
 Titan V & 12 GB  &  653 GB/s & 9.36 \\
 V100    & 16 GB  &  900 GB/s & 8.28 \\
 A100    & 40 GB  & 1500 GB/s & 6.47 \\
 \end{tabular}
\caption{List of the different GPUs used in this work and their specification.}
\label{tab:GPUs}
\end{table}

Thus, to measure the performance of our implementation, we will mostly use the achieved memory bandwidth as a metric.  This enables us to compare the achieved performance directly with what the hardware can theoretically support. The achieved bandwidth can be easily translated back into runtime, if desired. For the first order time splitting method, the domain has to be read and written seven times to perform the advections and to treat the source terms. Moreover, two additional reads are required to compute the charge density, which implies that the density function has to be accessed 16 times in total. Consequently, the amount of memory processed per time step is $16\cdot\text{dof}\cdot8$ bytes, where dof stands for the degrees of freedom and depends on the number of cells used, i.e., 
\[
\text{dof} = \prod_{i=1}^4 (\text{number of cells in dimension $i$})\cdot(k_i + 1),
\] 
where $k_i$ is the degree of the polynomial used in dimension $i$; in our case, $k_i=1$.
The factor 8 reflects the fact that the computations are performed in double precision; i.e. \texttt{sizeof(double)=8}. Then, the achieved bandwidth of our implementation is obtained by the following formula,
\[
\text{achieved bandwidth}=\frac{16\cdot\text{dof}\cdot8\cdot10^{-9} \text{ GB}}{\text{time per step in seconds}}.
\] 
Note that the time per step in seconds includes the computation of the electric field, while it is not included in the amount of memory accesses. Solving the quasi neutrality equation and the necessary interpolations and derivatives is a 3d problem, thus its impact on the full 4d drift-kinetic equation is small and not taking it into account does not change the overall bandwidth appreciably. Nevertheless, let us remark that its impact on the overall performance is significantly larger than solving the Poisson problem in the Vlasov--Poisson system in~\cite{Einkemmer2022}. 

The implementation of solving the 1d advections and the source term operates relatively close to the theoretical limit of the hardware and has thus a positive impact on the overall performance. For example, on the A100 GPU with a grid size of $64\times64\times64\times128$ cells, the advection in the $v$ dimension achieves 760~GB/s. Such a performance can not be expected for the 2d advection as the memory access pattern is more complicated and less predictable. Our implementation of the 2d advection achieves a bandwidth of~209~GB/s. 

The proposed second order splitting scheme achieves a bandwidth of~560~GB/s, while the first order scheme achieves~590~GB/s. This reduction is mainly due to the additional 2d advection that has to be performed. Overall this results in an increase in the run time of approximately~65\% for the second order scheme in contrast to the first order scheme. The second order scheme also requires~1.5~times more memory than the first order scheme since the density function has to be stored three times, compare algorithm~\ref{alg:splitting_1} to algorithm~\ref{alg:splitting_2}. However, it should be highlighted that the second order scheme (for the time step size chosen here) is more than an order of magnitude more accurate than the first order scheme (see figure \ref{fig:order_in_time}).

\begin{figure}[H]
 \centering
 \includegraphics[width=0.8\textwidth]{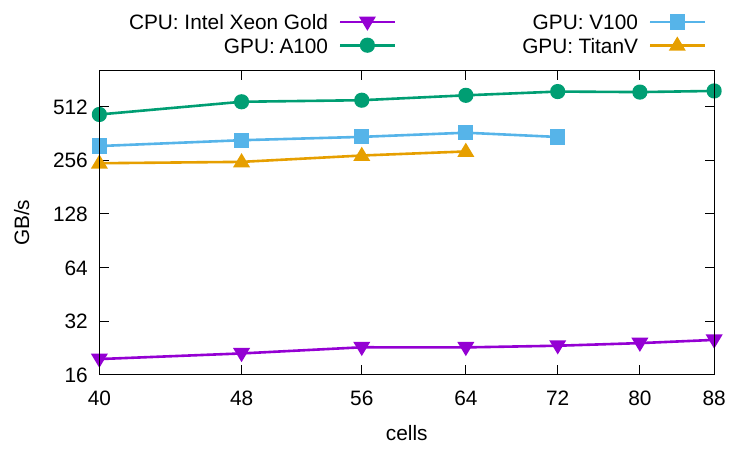}
 \caption{Achieved global bandwidth as a function of the number of cells used in each spatial dimension. In the velocity dimension, twice as many cells are used.}
 \label{fig:global-bandwidth}
\end{figure}

In figure~\ref{fig:global-bandwidth} the achieved bandwidth of the first order method is plotted as a function of the number of cells. For the A100 we observe a performance of approximately 590~GB/s for the algorithm. The A100 has a theoretical memory bandwidth of~1500~GB/s which corresponds to a parallel efficiency of approximately 40\% for our implementation. 

For the V100 we achieve a bandwidth of approximately 360 GB/s and on a Titan V 280~GB/s. Thus, the A100 is around 1.6~times faster than the V100 and 2.1~times faster than the Titan V and the improvement of performance between newer generation GPUs reflects exactly the improvement of bandwidth between them, which once more is a good indication that we are working on a memory bound problem.

Let us finally compare the GPU performance to the performance of a multi-core CPU system. We consider a dual socket system based on two Intel Xeon Gold 6226R CPUs. Our OpenMP implementation achieves a bandwidth of approximately 23 GB/s. It can be observed that all the GPUs outperform the CPU system by a significant margin, for example, the A100 is approximately a factor of 27~times faster than the CPU based system.

In order to better understand the achieved performance displayed in figure~\ref{fig:global-bandwidth}, we analyze the different parts of the splitting algorithm. The results are reported in table~\ref{tab:parts_gpu_vs_cpu}. Here we compare the performance between the A100 GPU and the CPU system where a fixed grid size of $64\times64\times64\times128$ cells is used. As expected, the GPU outperforms the CPU in each step. Moreover, as expected, the slowest part is the advection in the $r$-$\theta$ plane. However, while on the GPU this part is approximately a factor of four times slower than the other parts, on the CPU this gap is significantly smaller. Thus, as we might expect, the CPU copes better with the less predictable memory access of the 2d advection, even though overall the GPU in this step is still faster by approximately a factor of $10$.

\begin{table}[H]
\centering
 \begin{tabular}{l|r|r|r}
& CPU & GPU & bandwidth GPU \\ \hline
advection $r$-$\theta$ & 420.4 & 41.06 & 209.2 GB/s \\
advection $v$ & 370.8 & 11.30 & 760.0 GB/s \\
advection $z$ & 367.8 & 8.98 & 956.3 GB/s \\
compute $\rho$ & 387.9 & 6.22 & 690.5 GB/s \\
quasi neutrality & 83.4 & 4.09 &  \\
source term & 88.9 & 6.66 & 1290.1 GB/s \\
 \end{tabular}
 \caption{Average time per step of the different parts of the splitting algorithm in milliseconds and achieved bandwidth of these parts on the GPU. A grid size of $64\times64\times64\times128$ cells on an A100 GPU and a dual socket Intel Xeon Gold 6226R is used.}
 \label{tab:parts_gpu_vs_cpu}
\end{table}

To analyze the performance of the 2d SLDG algorithm in more detail, we report the timings of the required substeps in table~\ref{tab:sldg2d_gpu_vs_cpu}. As expected, the GPU is not significantly faster in computing the coefficients~\eqref{eq:sldg2d_coeff} (steps 1-5 in the table) as explained above. Nevertheless, the final step~\eqref{eq:sldg2d_implement} (step 6 in the table) on the GPU gives a significant boost in performance compared to the CPUs. 
As expected, computing the coefficients~\eqref{eq:sldg2d_coeff} in steps 1-5 are less expensive than the computation of~\eqref{eq:sldg2d_implement} implemented in step 6. These coefficients only depend on the electric field which is defined in the 3d spatial domain. In contrast, step 6 is responsible for a 2d advection in 4d phase space.
Let us also note that we do not observe a significant difference in the performance characteristics using different mesh sizes. Additionally, we do not expect an impact on performance using different time step sizes. The step size influences only the shape of the upstream cell, and since the current implementation allows the upstream cell to be a subset of $3\times3$ Eulerian cells, a comparable amount of data has to be processed to perform steps 1-5 in table~\ref{tab:sldg2d_gpu_vs_cpu}. The performance of step 6 is dictated by reading and writing the 4d domain, where the impact on the time step size is again rather small.

\begin{table}[H]
\centering
 \begin{tabular}{l*{2}{|>{\raggedleft\arraybackslash}p{4em}}*{2}{|>{\raggedleft\arraybackslash}p{4em}}}
mesh size & \multicolumn{2}{c|}{$64\times64\times64\times128$} & \multicolumn{2}{c}{$80\times80\times80\times160$} \\ \hline
& CPU & GPU & CPU & GPU \\ \hline
1. compute the upstream mesh & 3.61 & 0.43 & 6.00 & 0.81\\
2. get outer segments & 4.87 & 1.25 & 7.74 & 2.47 \\
3. get inner segments & 3.31 & 3.12 & 5.96 & 5.62 \\
4. solve adjoint problem & 3.04 & 0.43 & 5.55 & 0.78\\
5. compute integrals & 29.06 & 13.78 & 62.76 & 27.76 \\
6. finalize & 376.28 & 22.05 & 869.23 & 59.66 \\ \hline
total & 420.36 & 41.06 & 957.55 & 97.10 \\
 \end{tabular}
 \caption{Average time per step of the parts of the 2d SLDG algorithm in milliseconds. An A100 GPU and a dual socket Intel Xeon Gold 6226R is used.}
 \label{tab:sldg2d_gpu_vs_cpu}
\end{table}

\section{Acknowledgements}
This project has received funding from the European Union’s
Horizon 2020 research and innovation programme under
the Marie Skłodowska-Curie grant agreement No 847476.
The views and opinions expressed herein do not necessarily
reflect those of the European Commission.

\bibliographystyle{plain}
\bibliography{literatur}
\end{document}